\documentstyle[epsf%
,preprint%
,prl,aps]{revtex}

\firstfigfalse

\def\0{\over } \def\2{\textstyle{1\over2}} \def\4{\textstyle{1\over4}}
\def\5{\hat } \def\6{\partial }
 
\def\a{\alpha } \def\b{\beta }  \def\d{\delta }
\def\e{\epsilon }

 \def\o{\omega } 
\def\ph{\varphi }
\def\mn{{\mu\nu}}

\def\({\left(} \def\){\right)} \def\<{\langle } \def\>{\rangle }
\def\[{\left[} \def\]{\right]}  

\newcommand{\bea}{\begin{eqnarray}}
\newcommand{\eea}{\end{eqnarray}}
\newcommand{\be}{\begin{equation}}
\newcommand{\ee}{\end{equation}}
\newcommand{\nn}{\nonumber\\ }

\def\Tr{{\,\mathrm Tr\,}}
\def\Im{{\,\mathrm Im\,}}
\def\Re{{\,\mathrm Re\,}}

\begin{document}
\preprint{CERN-TH/99-184, SACLAY-T99/055, TUW-99/13}
%\tighten
\draft
%\wideabs{
\title{The entropy of the  QCD plasma}
\author{J.-P. Blaizot}
\address{Service de Physique Th\'eorique, CE Saclay,
	F-91191 Gif-sur-Yvette, France}
\author{E. Iancu}
\address{Theory Division, CERN, CH-1211 Geneva 23, Switzerland}
\author{A. Rebhan}
\address{Institut f\"ur Theoretische Physik,
         Technische Universit\"at Wien,\\
         Wiedner Hauptstra\ss e 8-10/136,
         A-1040 Vienna, Austria}
\date{June 11, 1999}
\maketitle
\begin{abstract}
Self-consistent approximations in
terms of fully dressed propagators provide a simple expression for the entropy
of an ultrarelativistic plasma, which isolates
the contribution of the elementary excitations as a leading contribution. 
Further approximations, whose validity is checked 
on a soluble model involving a
scalar field, allow us to calculate  the entropy of the QCD plasma. 
We obtain an accurate description of lattice data for purely gluonic QCD, 
down to temperatures of about twice the transition temperature.    
\end{abstract}

%}

The properties of the high temperature phase of QCD
cannot be easily calculated
using perturbation theory, in spite of the fact that the gauge coupling $g$ is
small if the temperature $T$ is sufficiently high. This is evidenced in
particular by  the poor convergence properties of the perturbative series 
\cite{QCDP}. 

Lattice results, which  show  that the ideal gas
limit is approached as $T$ becomes large, 
can be accounted for reasonably well by
phenomenological fits involving  massive quasi-particles \cite{Peshier,LH}. 
Although the
quasiparticle picture suggested by such fits is 
a rather crude representation of
the actual physics of non-abelian gauge theories, it supports 
the idea that one
should be able to give an accurate description of the
thermodynamics of the QCD plasma in terms of its elementary excitations.

It is worth emphasizing at this stage that, among the relevant degrees of 
freedom,  the soft collective ones, with momenta of order $gT$, are clearly 
non-perturbative. 
Although their leading order  contribution $\propto g^3$ 
to the pressure can be
easily isolated \cite{Kapusta}, it does not  make  
much physical  sense to
regard this contribution as a genuine perturbative
correction. Recent investigations \cite{Pade,KPP,DHLR2} 
indicate indeed that trying to
represent this contribution by a 
truncated polynomial in $g$ is not appropriate.

In order to carry out a more complete calculation, we shall  use techniques
which allow systematic rearrangements of the perturbative expansion, 
avoiding double
countings.  We shall rely in particular 
on self-consistent approximations which provide
a simple expression for the entropy, isolating the
contribution of the elementary excitations as a leading contribution. 
We show  that this
entropy formula  can be used to get a good estimate of the QCD entropy at high
temperature. The results that we have obtained 
so far are quite encouraging and give us
the hope that an analytical control of the high temperature phase of QCD is 
within reach. 

We shall first discuss scalar field theories with $g\ph^3+g^2\ph^4$
interactions, with the double purpose of 
presenting the general framework and of 
checking  approximations which will be used later for QCD.

The thermodynamic potential $\Omega=-PV$ 
of the  scalar field can be written as the
following  functional of the full propagator $D$ \cite{LW,Baym}:
\be
\label{LW}
\b \Omega[D]=-\log Z=\2 \Tr \log D^{-1}-\2 \Tr \Pi D+\Phi[D]
\ee
where $\Tr$ denotes the trace in configuration space,
$\b=1/T$, and
$\Phi[D]$ is the sum of the 2-particle-irreducible ``skeleton''
diagrams
\be\label{skeleton}
-\Phi[D]= 
\epsfxsize=7cm
\epsfbox[50 390 550 430]{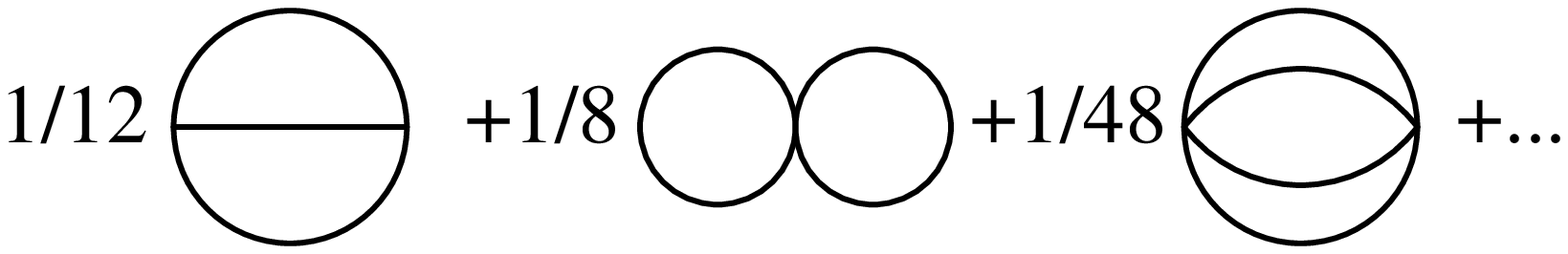}
\ee
The self energy $\Pi=D^{-1}-D^{-1}_0$, 
where $D_0$ is the bare propagator, is
related to $\Phi[D]$ by 
\be\label{PhiPi}
\d\Phi[D]/\d D=\2\Pi.
\label{Pi}
\ee
An important property of the functional  $\Omega[D]$, which is easily verified
using  (\ref{PhiPi}), is that it is stationary under variations of
$D$:
\be\label{selfcons}
{\d\Omega[D] / \d D}=0.
\ee
Self-consistent (``$\Phi$-derivable'')
\cite{Baym} approximations, i.e., approximations which preserve this property, 
are  obtained by selecting a class of skeletons in 
$\Phi[D]$ and calculating $\Pi$ from Eq.~(\ref{Pi})
above.

The stationarity of $\Omega[D]$ has an interesting consequence for  the
entropy. Because of  
Eq.~(\ref{selfcons}) the temperature derivative of the spectral
density in the dressed propagator cancels out in 
the entropy density
\be
{\cal S}=-{\6(\Omega/V)/\6T}
\ee 
and one obtains \cite{Riedel,VB}:
\bea\label{Ssc}
{\cal S}&=&-\int\!\!{d^4k\0(2\pi)^4}{\6n(\o)\0\6T} \Im \log D^{-1}(\o,k) \nn
&&+\int\!\!{d^4k\0(2\pi)^4}{\6n(\o)\0\6T} \Im\Pi(\o,k) \Re D(\o,k)+{\cal S}'
\eea
with
\be
{\cal S}'=-{\6(T\Phi)\0\6T}\Big|_D+
\int\!\!{d^4k\0(2\pi)^4}{\6n(\o)\0\6T} \Re\Pi \Im D=0
\ee
up to terms that are of loop-order 3 or higher.  Thus, in contrast to
$\Omega$, where
$\Phi$ contributes already to order
$g^2$ in perturbation theory, Eq.~(5) with ${\cal S}'=0$ is perturbatively
correct up to, and including, order $g^3$ \cite{us}.  The
first two terms in Eq.~(\ref{Ssc}) represent essentially the entropy of
``independent quasiparticles'', while ${\cal S}'$ may  be viewed as the
residual interactions among these quasiparticles \cite{VB}.

Besides this important simplification, ${\cal S}$, in contrast to
the pressure, has the
advantage of manifest ultra-violet finiteness, since ${\6n/\6T}$
vanishes exponentially for both $\o\to\pm\infty$.
Moreover, any multiplicative renormalization $D\to ZD$, $\Pi\to Z^{-1}\Pi$
with real $Z$ drops out from Eq.~(\ref{Ssc}). 

We now focus on the self-consistent approximation obtained in $g^2\varphi^4$
theory where 
only the second diagram in Eq.~(\ref{skeleton}) for $\Phi$ is kept.
Then  
$\Im\Pi=0$ and $\Re\Pi=m^2=const.$, and  Eq.~(\ref{Ssc}) reduces to
\bea\label{Sscphi4}
{\cal S}&=&-\int\!\!{d^4k\0(2\pi)^4}{\6n(\o)\0\6T}\Im\log (k^2-\o^2+m^2) \nn
&=&{4\0T}\left[{\pi^2T^4\090}-{m^2T^2\048}+{m^3T\048\pi}+\ldots \right].
\eea
A fully self-consistent determination of $m$ corresponds to solving the
gap equation
\be\label{gap}
m^2=12g^2\int\!\!{d^4k\0(2\pi)^3}n(\o)\e(\o)\d(\o^2-k^2-m^2).
\ee
When the solution of this equation is inserted in  Eq.~(\ref{Sscphi4}), the
entropy obtained coincides with that of the  exact solution of a scalar
O($N$)-model in the limit of
$N\to\infty$
\cite{DJ,DHLR2}. Note that in contrast to
Eq.~(\ref{Sscphi4}), the gap equation  is ultraviolet divergent (for
$\o\to-\infty$),  and requires renormalization
\cite{DHLR2},
affecting $m^2$ at perturbative order $g^4$ and beyond. 

In view of the subsequent application to QCD, where a  fully
self-consistent determination of the gluonic self-energy  seems
prohibitively difficult, we consider now  perturbative approximations. Our goal
is to  obtain approximate expressions for the self energy which allow us to
reproduce the perturbative result for the 
entropy when expanded to order $g^3$. We
emphasize that our final results for the entropy are non-perturbative, and not
limited to a truncated polynomial in 
$g$. What we are testing here is  the quality of approximations which preserve
self-consistency up to order $g^3$ at least, 
which is what we shall be able to do
in QCD. 

As first approximation we consider the leading
contribution to the self-energy at high temperature, 
the so-called hard thermal loop (HTL) \cite{BP}, and refer to Eq.~(\ref{Ssc})
with this restriction as ${\cal S}_{\rm HTL}$.
For the $\varphi^4$ theory, we have simply
$\Pi\to\hat\Pi=\hat m^2=g^2T^2$. 
When inserted into Eq.~(\ref{Sscphi4}), this
yields the correct result for the leading-order 
interaction term $g^2T^3$ in the
entropy. 

On the other hand, the order $g^3$ contribution contained in 
${\cal S}_{\rm HTL}$
%following from Eq.~(\ref{Sscphi4}) 
turns out to be too small by a factor of 4 
when compared to the well-know perturbative result\cite{Kapusta}.
This is corrected by including the next-to-leading order (NLO) term in
the thermal mass through resummed perturbation theory, 
$m^2=\5m^2+\d m^2=g^2T^2-{3\0\pi}g^3T^2$ \cite{Kapusta}.
So, when compared to conventional resummed perturbation theory, the
order-$g^3$ term arises in an unusual manner: whereas in the former
the entire plasmon effect comes from the infrared regime,
in Eq.~(\ref{Ssc}) %this region contributes only part, 
an even larger
contribution comes indirectly from the infrared through corrections
to the dispersion laws relevant at hard momenta (the $T^2$ term in 
Eq.~(\ref{Sscphi4}) comes entirely from hard $k\sim T$). This may be understood
as a consequence of the requirement of self-consistency: recall that 
 Eq.~(\ref{Sscphi4})  relies on the stationarity of the thermodynamic potential,
and this has to be maintained at the order of interest. 

At large coupling the NLO result
for
$m^2$ inevitably turns negative. This can be avoided by taking instead the
perturbatively equivalent form $\hat m^2+\d m^2=g^2T^2/[ 1+3g/\pi]$, which
is monotonous in $g$, giving 
a very good approximation to the solutions of Eq.~(\ref{gap}) up to
 $g\gtrsim 1$.

In Fig.~1 we compare the various approximations numerically with
the exact  entropy (full line),  normalized to their ideal-gas values (SB), 
as functions of the renormalized coupling
in the $\overline{\hbox{MS}}$-scheme.
In contrast to the full result, the perturbative approximations are
renormalization-scale dependent. As in Ref.~\cite{DHLR2} we
consider the effect of varying the scale $\bar\mu$, here in the
range $\bar\mu=(\2\ldots 2)\times 2\pi T$. 
The lower and upper
dark-gray bands correspond to conventional perturbative results 
for $S/S_{\rm SB}$
up to order $g^2$ and $g^3$, resp.,
the medium-gray bands to the result of HTL- and NLO-resummations
of the two-loop entropy. The latter, which clearly represent a 
substantial improvement over the former, are 
the approximations that we shall now
implement for QCD.  

%%%%%%%%%%%%%%%%%%%%%%%%%%%%%%%%%%%%%%%%%%%%%%%%%%%%%%%%%%%%%%%%%%%%
\begin{figure}
%\vspace{-4mm}
\epsfxsize=10cm
\centerline{\epsfbox[70 210 540 520]{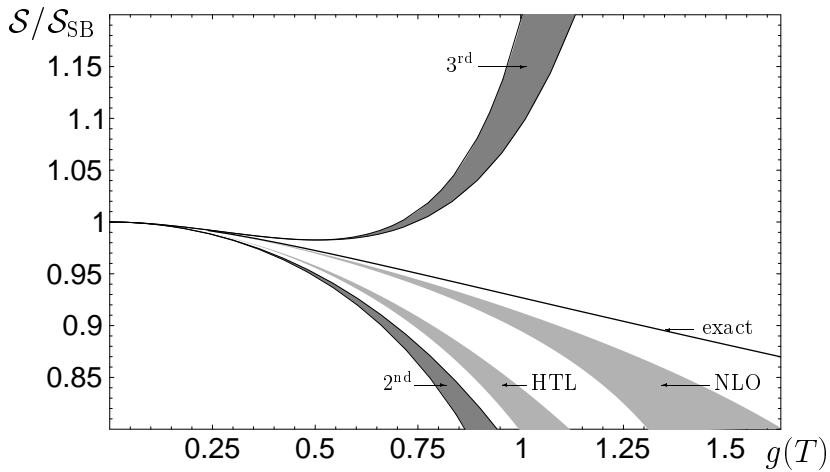}}
%\vskip4mm
\caption{Comparison of perturbative and HTL-improved approximations to
the entropy in the large-$N$ scalar O($N$)-model.
See text for detailed explanations.}
\label{fig1}
\end{figure}
%%%%%%%%%%%%%%%%%%%%%%%%%%%%%%%%%%%%%%%%%%%%%%%%%%%%%%%%%%%%%%%%%%%

The analog of Eq.~(\ref{Ssc}) in purely gluonic QCD 
and in Coulomb gauge
reads
\bea\label{SQCD}
{\cal S}&=& -N_g\int\!\!{d^4k\0(2\pi)^4}{\6n(\o)\0\6T}
\Bigl\{2\Im\log(-\o^2+k^2+\Pi_T)\nn&&\qquad\qquad\qquad\quad
+2\Im\Pi_T\Re[\o^2-k^2-\Pi_T]^{-1}\nn&&\quad
+\Im\log(k^2+\Pi_L)-\Im\Pi_L \Re[k^2+\Pi_L]^{-1}\Bigr\}
\eea
with $N_g=N^2-1=8$ for SU(3). The (spatially)
 transverse $(\Pi_T)$ and longitudinal $(\Pi_L)$ structure functions will be
specified below.  
Here we have assumed that the gluon self-energy is transverse
with respect to the four-momentum,
and that there are no contributions from
the ghosts, which turns out to be justified
in the approximations that we shall be interested in \cite{us}.

The order $g^2$ contribution to the entropy is easily extracted from
Eq.~(\ref{SQCD}):
\bea\label{S2}
{\cal S}^{(2)} &=&-2\pi N_g\int\!\!{d^4k\0(2\pi)^4}{\6n\0\6T} 
\e(\o)\d(\o^2-k^2)
\Re\Pi_T(\o,k)
\nn
&=&-N_g{m_\infty^2T\06}=-{NN_g\036}g^2T^3.
\eea
Here we have used the fact that the integral is dominated
by hard momenta and that the transverse quasiparticles have the
asymptotic thermal mass:
\be\label{minfty}
m_\infty^2=\Pi_T(\o^2=k^2)=g^2NT^2/6.
\ee
This latter result is exact at the bare one-loop level  \cite{KKR}.

The contribution of order $g^3$ involves 
loop integrals with soft momenta, which
requires using the HTL approximation for $\Pi$, where
\cite{KKW,BP}: 
\bea
\label{PiL}
\5\Pi_L(\o,k)&=&\textstyle \5m_D^2[1-{\o\02k}\log{\o+k\0\o-k}],\\
\label{PiT}
\5\Pi_T(\o,k)&=&\2[\5m_D^2+({\o^2\0k^2}-1)\5\Pi_L],
\eea
with $\5m_D=gT\sqrt{N/3}$.
The spectral density of the corresponding 
 gluon propagator consists of quasiparticle poles with
momentum-dependent effective masses and Landau damping cuts for $|\o|<k$.
When  $k\gg gT$, the additional pole 
associated to the collective longitudinal excitation has exponentially
vanishing residue
\cite{P}.

The order-$g^3$ contribution in ${\cal S}_{\rm HTL}$ can be isolated as
\bea
&&{\cal S}^{(3)}_{\mathrm HTL}/N_g = \int\!\!{d^4k\0(2\pi)^4}{1\0\o} \Bigl\{
2\Re\5\Pi_T \Bigl[ \Im[\o^2-k^2-\5\Pi_T]^{-1} \nn
&& -\Im[\o^2-k^2]^{-1} \Bigr] 
-\Re\5\Pi_L \Im[k^2+\5\Pi_L]^{-1} \Bigr\} -{\hat m_D^3\024\pi}
\label{S3a}
\eea
For the same reason as in the above scalar example,
this is only part of the full order $g^3$ contribution. 
Remarkably, it turns out that, as in the scalar case, Eq.~(\ref{S3a}) is 
precisely 1/4 of the
correct result $S^{(3)}=N_g \hat m_D^3/(3\pi)$.

%%%%%%%%%%%%%%%%%%%%%%%%%%%%%%%%%%%%%%%%%%%%%%%%%%%%%%%%%%%%%%%%%%%%
\begin{figure}
%\vspace{-4mm}
\epsfxsize=8cm
\centerline{\epsfbox[60 400 300 460]{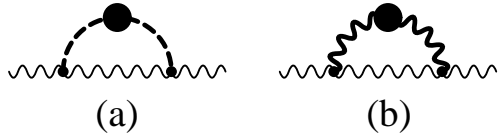}}
%\vskip2cm
\caption{NLO contributions to $\d\Pi_T$ at hard momentum. Thick
dashed and wiggly lines with a blob represent HTL-resummed longitudinal
and transverse propagators.%, resp.
}
\label{fig2}
\end{figure}
%%%%%%%%%%%%%%%%%%%%%%%%%%%%%%%%%%%%%%%%%%%%%%%%%%%%%%%%%%%%%%%%%%%

The remaining order-$g^3$ correction  comes from Eq.~(\ref{S2}) with
$\d \Pi_T$ in place of $\5\Pi_T$,   $\d \Pi_\mn$ being evaluated at order
$g\5m_D^2$ in HTL-resummed perturbation theory. The  expression (\ref{S2}) is
dominated by hard momenta $k\sim T$, and  to the order of interest $\d \Pi_T$ is
given by the two contributions shown in Fig.~2, in which one internal line
is hard,  and the other one is a soft
resummed longitudinal (a) or transverse (b) gluon propagator.
Diagram (a)  restores the
correct combinatorial factor of the longitudinal ring diagrams,
whereas diagram (b) compensates for spurious transverse plasmon effects
that are present in the HTL approximation ${\cal S}_{\rm HTL}$
\cite{us}. 
  
We turn now to the  numerical evaluation of ${\cal S}_{\rm HTL}$
and shall discuss the effects of the above NLO contributions further
below. ${\cal S}_{\rm HTL}$ involves
two physically distinct contributions. One corresponds to the transverse
and longitudinal gluonic quasiparticle poles,
\bea\label{SQP}
{\cal S}^{\mathrm QP}_{\mathrm HTL}&=&- N_g\int\!\!{d^3k\0(2\pi)^3} {\6\0\6T}
\Bigl[2T\log(1-e^{-\o_T(k) / T}) \nn
&&\qquad\qquad+ T\log{1-e^{-\o_L(k) / T}\01-e^{-k/T}} \Bigr],
\eea
where only the explicit $T$ dependences are to be differentiated,
and not those implicit in the HTL dispersion laws $\o_T(k)$ and $\o_L(k)$.
The latter are given by the solutions of $\o_T^2-k^2=\5\Pi_T(\o_T,k)$
and $k^2=-\5\Pi_L(\o_L,k)$ \cite{KKW}.

Secondly, there are the Landau-damping contributions which read
\bea\label{SLD}
{\cal S}^{\mathrm LD}_{\mathrm HTL}&=&
-N_g \int\limits_0^\infty{ k^2 dk \0 2\pi^3} 
\int\limits_0^k \! d\o {\6n(\o)\0\6T} 
\Bigl\{ 2 \arg [ k^2-\o^2+\5\Pi_T ] \nn
&&\qquad\qquad -2\Im \5\Pi_T \Re[\o^2-k^2-\5\Pi_T]^{-1} \nn
&&+ \arg [ k^2+\5\Pi_L ] - \Im \5\Pi_L \Re[k^2+\5\Pi_L]^{-1} \Bigr\}.
\eea

The usual perturbative $g^2$-contribution (\ref{S2}) is contained in the first
term of Eq.~(\ref{SQP}); all the other terms in Eqs.~(\ref{SQP}),(\ref{SLD})
are of order $g^3$ in a small-$g$ expansion.

In Fig.~3, we compare the numerical evaluation of ${\cal S}_{\rm HTL}/
{\cal S}_{\rm SB}$
with the lattice data for purely gluonic QCD from Ref.~\cite{Boyd}, in the
same manner as done in Ref.~\cite{ABS}, i.e., we use the two-loop
running coupling constant $\a_s(\bar\mu)$ 
of the $\overline{\mbox{MS}}$-scheme
with $\Lambda_{\overline{\mbox{MS}}}=T_c/1.03$ and the renormalization
scale is varied in the range $\bar\mu=(\2\ldots 2)\times 2\pi T$
to give an estimate of the theoretical uncertainty. 

The thick dark-gray line represents the lattice data for the entropy density
with the thickness of the line giving roughly the error reported in
Ref.~\cite{Boyd}. Our result reproduces the lattice data rather well
already for $T\gtrsim 2T_c$. 

%%%%%%%%%%%%%%%%%%%%%%%%%%%%%%%%%%%%%%%%%%%%%%%%%%%%%%%%%%%%%%%%%%%%
\begin{figure}
%\vspace{-4mm}
\epsfxsize=10cm
\centerline{\epsfbox[70 200 540 510]{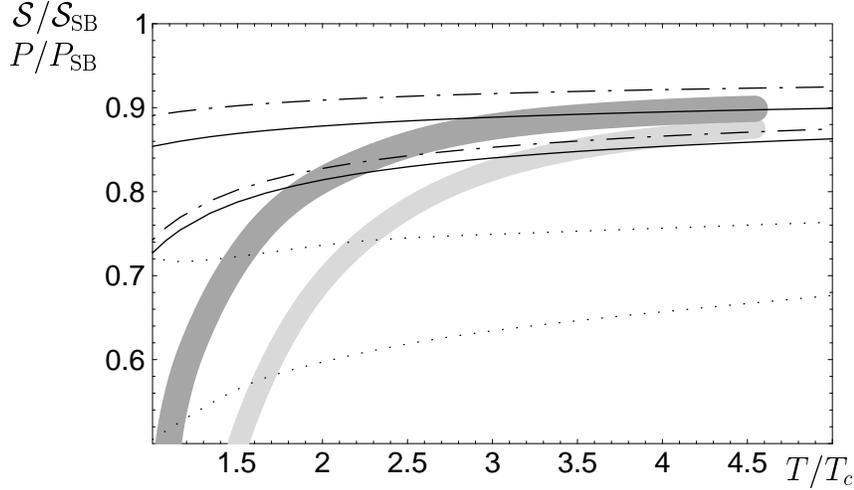}}
%\vskip4mm
\caption{
HTL-improved results for the 2-loop 
entropy ${\cal S}/{\cal S}_{\rm SB}$ in purely
gluonic QCD (full lines)
with $\bar\mu$ varied between $\pi T$ and $4\pi T$;
our estimates for NLO effects are given by the dash-dotted lines.
The lattice result
for the entropy
is represented by the dark-gray band. For comparison, 
the HTL-resummed
results of Ref.~\protect\cite{ABS} 
for the 1-loop pressure are given by the
dotted lines, the lattice results for $P/P_{\rm SB}$ by the
light-gray band. [Because of the weak temperature-dependence of the
theoretical results the predictions for ${\cal S}/{\cal S}_{\rm SB}$ are %to
%a good approximation 
approximately
those for $P/P_{\rm SB}$ and vice versa.]
}
\label{fig3}
\end{figure}
%%%%%%%%%%%%%%%%%%%%%%%%%%%%%%%%%%%%%%%%%%%%%%%%%%%%%%%%%%%%%%%%%%%

Also given in the same figure are the lattice data for the pressure
$P/P_{\rm SB}$ (the lower light-gray band) and the result 
for the full HTL-resummation of the
one-loop pressure reported in Ref.~\cite{ABS} (dotted lines). 
The discrepancy between
the latter
can be attributed in part to an overcounting in the interaction
pressure in Ref.~\cite{ABS}, which will be corrected only in
a fully resummed two-loop calculation. 
By contrast, our approach has the advantage of including the correct leading
order interaction terms already in a pure HTL approximation, 
and also of manifest
ultra-violet finiteness, thus avoiding the introduction of artificial 
counter-terms depending on the thermal mass. 

Turning now to the NLO approximation to ${\cal S}$, 
we note that, at hard momenta
$k$,
$\Re\d\Pi_T(\o^2=k^2)$   is, unlike
$m_\infty^2$, a nonlocal
quantity.  
In order to get some estimate on its numerical
effect, we approximate it by a constant correction which has
the right magnitude to produce the known perturbative coefficient
of order $g^3$.
We choose this as $m_\infty^2+\d m_\infty^2=g^2T^2N/(6[ 1+\sqrt{3N}g/\pi])$,
and include this correction
only in the hard momentum region defined by
$k>M=\sqrt{2\pi T m_D}$. The reason for this is that we do not want
to change by hand the overall scale of HTL contributions in the soft
regime, where NLO contributions are known to behave quite differently: 
the long-wavelength plasma frequency receives much smaller negative corrections 
\cite{Sch}, and the Debye screening
mass is known to be even substantially
increased %by NLO corrections
\cite{RDeb,LP}.

The dash-dotted lines in Fig.~3 give 
the correspondingly modified numerical results.
In addition to a variation of $\bar\mu$ we have also included a variation
of $M$, the boundary between hard and soft momenta, by a factor of 2
around its central value.
These results happen to describe the lattice data surprisingly well, although
their primary significance is to demonstrate the relative stability
of our scheme upon inclusion of terms that restore
equivalence with the known perturbative result up to and
including order $g^3$.
A full NLO calculation still remains to be done and will be presented in a
forthcoming publication.

This work was supported by the Austrian-French scientific exchange
program Amadeus of APAPE and \"OAD.

\end{document}